\documentclass[prb,twocolumn,superscriptaddress,floatfix,citeautoscript]{revtex4-2}

\usepackage{graphicx} 
\usepackage{amsmath,bm}	
\usepackage{amsmath,amssymb}
\usepackage{graphicx,epsfig,sidecap,wasysym}
\usepackage{xcolor,epstopdf}
\usepackage[colorlinks,bookmarks=false,citecolor=blue,linkcolor=blue,urlcolor=blue]{hyperref}
\usepackage{soul}

\newcommand{\be}{\begin{equation}}
\newcommand{\ee}{\end{equation}}

\begin{document}
\title{ Many-body localization transition in a frustrated XY chain} 

 
\author{M.\,S. Bahovadinov}
\affiliation{Russian Quantum Center, Skolkovo, Moscow 143025, Russia}
\affiliation{  Physics Department, National Research University Higher School of Economics, Moscow, 101000, Russia}
\author{D.\,V. Kurlov}
\affiliation{Russian Quantum Center, Skolkovo, Moscow 143025, Russia}
\author{S.\,I. Matveenko}
\affiliation{Russian Quantum Center, Skolkovo, Moscow 143025, Russia}
\affiliation{L. D. Landau Institute for Theoretical Physics, Chernogolovka, Moscow region 142432, Russia }
\author{B.\,L. Altshuler}
\affiliation{Physics Department, Columbia University, 538 West 120th Street, New York, New York 10027, USA}
\affiliation{Russian Quantum Center, Skolkovo, Moscow 143025, Russia}
 \author{G.\,V. Shlyapnikov}
\email{shlyapn@lptms.u-psud.fr}
\affiliation{Russian Quantum Center, Skolkovo, Moscow 143025, Russia}
\affiliation{Moscow Institute of Physics and Technology, Inst. Lane 9, Dolgoprudny, Moscow Region 141701, Russia}
\affiliation{Universit\'e Paris-Saclay, CNRS, LPTMS, 91405 Orsay, France}
\affiliation{Van der Waals-Zeeman Institute, Institute of Physics, University of Amsterdam,Science Park 904, 1098 XH Amsterdam, The Netherlands}


\begin{abstract}
We demonstrate many-body localization (MBL) transition in a one-dimensional isotropic XY chain with a weak next-nearest-neighbor frustration in a random magnetic field. We perform finite-size exact diagonalization calculations of level-spacing statistics and fractal dimensions to characterize the MBL transition with increasing the random field amplitude. An equivalent representation of the model in terms of spinless fermions explains the presence of the delocalized phase by the appearance of an effective non-local interaction between the fermions. This interaction appears due to frustration provided by the next-nearest-neighbor hopping.  
\end{abstract}

\maketitle


\section{Introduction}
 The interplay of interparticle interactions and disorder in low-dimensional quantum systems is an old problem~\cite{Apel,Giamarchi} and it has been an active research direction in the recent years. The majority of these studies have been dedicated to many-body localization~\cite{Altshuler}, the phenomenon that extends Anderson localization (AL) to interacting many-particle systems. In particular, numerical studies of disordered interacting one-dimensional (1D) quantum systems demonstrate a transition to the many-body localized (MBL) phase, if the amplitude of diagonal disorder is sufficiently large (for recent reviews see Refs.~\cite{ReviewLuitz,ReviewAbanin} ). At weak disorder, the system is in the ergodic phase and the eigenstate-thermalization hypothesis (ETH)~\cite{Deutch, Srednicki} is obeyed. On the contrary, in the MBL phase the ETH is violated~\cite{Pal_Huse_2010,YBLev2014,Serbyn2015,Luitz2016,De_Luca_2013, De_Luca_2014}, which implies protection of quantum states from decoherence and opens new prospects for quantum information storage. Recent studies of the MBL phase also show the area law entanglement entropy ~\cite{Bardarson,Nayak_2013,Serbyn2013}, Gumbel statistics for entanglement spectra~\cite{Gritsev2019,Serbyn2016} and vanishing steady transport~\cite{ Berkelbach2010,Barisic2016,YBLev2015,Herbrych2017}. A number of these properties can be explained in terms of emerging quasi-local integrals of motion and the resulting quasi-integrability~\cite{Papic2013,Rademaker2016,Chandran2015}. The MBL transition is usually characterized by the level-spacing statistics~\cite{KudoRn,LuitzAletRn,OganesyanHuseRn,CorentinRn, ShengRn,KhemaniRn}, participation entropies~\cite{AletDq,SantosDq}, underlying entanglement structure of eigenstates~\cite{GrayEnt,ShengRn,Gritsev2019}, occupation spectrum of the one-particle density matrix~\cite{OPDMMeissner,OPDMBuijsman,OPDMImura} and by quantum correlations of neighboring states~\cite{PinoDqKl,LuitzAletRn}. 

\begin{figure}[t]
\includegraphics[width=8 cm ]{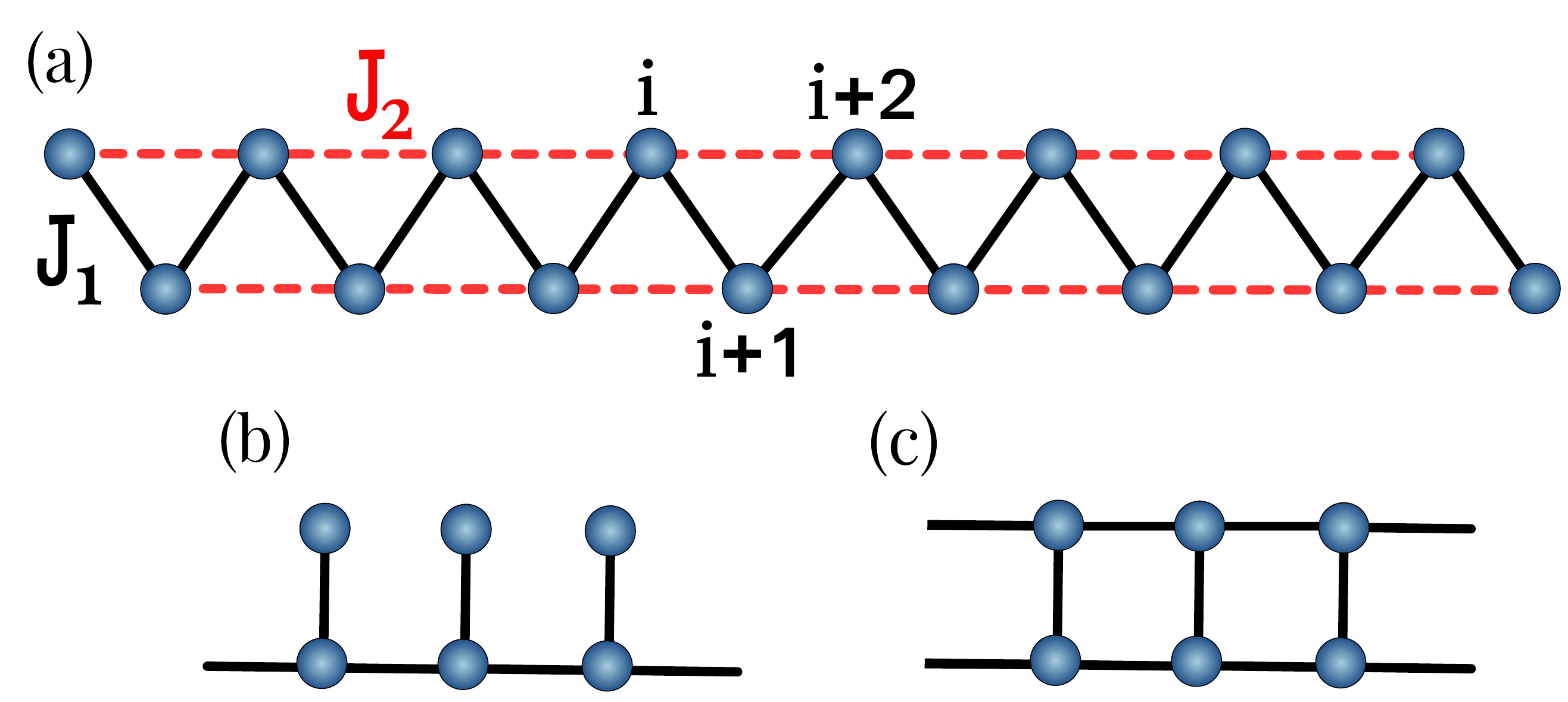}  
 \caption{\label{fig:fig0} 
   (a) The XY model~(\ref{eq:H}) is considered in a zig-zag ladder with the nearest-neighbour $J_1$ and next-nearest neighbour $J_2$ flip-flop amplitudes. The proposed MBL transition can be equivalently realized in the XY model considered in quasi-1D geometries such as (b) in the Kondo lattice or (c) in a two-leg ladder. 
}
\end{figure} 
When studying the MBL transition for trapped ions, polar molecules, or superconducting circuits one should take into account long-range interaction and hopping terms~\cite{ Anderson, BurinKagan, Saffman,Blatt,Pfau,Bohn,Monroe}. The recent proposal \cite{Proposal} suggests realization of the 1D XY model using superconducting arrays of three-dimensional transmons with interqubit dipolar interactions. In the proposed model only the nearest-neighbor and next-nearest-neighbor flip-flop amplitudes are present, whereas the other amplitudes exactly vanish (see Fig.~\ref{fig:fig0}(a)). It is reasonable to analyse the interplay between the frustration natural to XY magnets with long-range flip-flop amplitudes, and the diagonal disorder in this simplified model.
 
In the present paper we show the existence of the many-body localization - delocalization transition in the latter model. We demonstrate that the presence of the next-nearest-neighbor flip-flop term, which provides the frustration, is crucial for the MBL transition. This is revealed by using the Jordan-Wigner  transformation, which maps the original spin model  onto the model of spinless fermions. The fermionic model consists of trivial terms of hoppings to the two nearest sites and a non-local interaction term. In the absence of frustration in the original spin model, the non-local interaction and next-nearest-neighbor hopping terms vanish and we have the system of free fermions where the states are known to be localized in arbitrarily weak potential disorder~\cite{Mott,Klein}. A similar scenario occurs in disordered XY model considered in another quasi-1D geometries, such as in the Kondo lattice or in a two-leg ladder [Fig.~\ref{fig:fig0}(b)-(c)].
  
The outline of the paper is as follows. In Sec. II we present the model of our study and discuss the main symmetries and properties. We then represent the model in terms of interacting spinless fermions and provide qualitative discussion on the origin of the present interaction in Sec. III. In Sec. IV  we introduce localization measures we used to characterize the MBL transition and provide numerical confirmations of the predicted transition in Sec. V. Our outlook and concluding remarks are given in Sec. VI.

 
 

\label{sec:model}
 
\section{Model and symmetries}
We consider a one-dimensional frustrated spin-1/2 isotropic $XY$ spin chain~\cite{XY} in a random magnetic field. The Hamiltonian of the system has the following form: 
\begin{equation}
 H =  \sum_{\beta=1,2}J_\beta \sum_{i=1}^{L} \left[ S^x_i S^x_{i+\beta}+S^y_i S^y_{i+\beta} \right]+ \sum_{i=1}^L h_i S_i^z,
\label{eq:H}
\end{equation}
where $J_{1,2} > 0$ are exchange interaction coupling constants, and $h_j$ are uncorrelated random field amplitudes drawn from the uniform distribution $[-h,h]$.  We consider the model in a chain with an even number of sites~$L$ and impose periodic boundary conditions.  
 In the presence of disorder the Hamiltonian~(\ref{eq:H}) possesses only the $U(1)$ symmetry. If either $J_1 = 0$ or $J_2 = 0$, the eigenstates  are localized for arbitrary~$h_j$ since in this case the model can be mapped onto the system of free fermions. The discussion of integrability in this limit is presented in Appendix A.    
 
The low-temperature phase diagram of the clean model (\ref{eq:H}) was earlier studied extensively using numerical techniques such as density matrix renormalization group (DMRG) and Exact Diagonalization (ED)  methods~\cite{DMRGXY,EDXY}. It is shown that for $ \kappa=J_2/J_1 \lesssim 0.32 $ the system is in the gapless Tomonaga-Luttinger liquid phase, where expectation values of spin operators vanish and two-point correlation functions exhibit power-low decay.  
For  $ \kappa  \gtrsim 0.32 $ the gapped insulating phase (singlet dimer phase) is developed via the  Berezinskii-Kosterlitz-Thouless-type transition. In the limits of $\kappa=0$ and $\kappa=\infty$ the clean model~(\ref{eq:H}) is integrable. In the former case, the unfrustrated $XY$ chain is restored, while the second case corresponds to two decoupled $XY$ chains. For $\kappa \ll 1$ the  clean model is {\it quasi-integrable}, possessing {\it quasi-conserved} charges, that is charges conserved on a large time scale $\sim  \kappa^{-2}$ (see Appendix B and   Ref.~\cite{Kurlov2021}   for details). We hereafter consider only $ \kappa < 0.32 $, so that the system is in the gapless phase.  
  \section{Qualitative arguments for MBL}
\textit{Jordan-Wigner fermionization.} Before turning to our numerical results we point out that there is an implicit interaction which guarantees delocalization at weak disorder.  
 Naively, one may expect that spin fluctuations in our model should be localized by disorder like fermionic fluctuations, and the states should undergo AL in arbitrarily weak random magnetic fields. We show here that this is not the case. This is due to the statistics of spins which has neither pure bosonic nor pure fermionic character. However, one may fully fermionize the spin problem (\ref{eq:H}) by using the Jordan-Wigner (JW) transformation~\cite{JWT}:
\begin{equation}  \label{eq:JWTr}
S^{+}_i=c^{\dagger}_ie^{i \pi \sum_{p<i}\hat{n}_p}, \qquad 
S^{z}_i=\hat{n}_i -1/2,
\end{equation}
where $\hat{n}_i=c_i^{\dagger}c_i$ and  the operators $c^\dagger_i, c_i$ obey the canonical fermionic anti-commutation relations. 
From~Eq.~(\ref{eq:JWTr}) one immediately gets $S^{+}_iS^{-}_{j}=c^\dagger_i \hat{\Phi}_{i,j} c_j$, where the Hermitian operator $\hat{\Phi}_{i,j}$ is given by
\begin{equation} \label{eq:Phase}
\hat{\Phi}_{i,j} =\prod_{l = i + 1}^{j-1}(1-2\hat{n}_l), \qquad   j \geq i + 2, 
\end{equation}
and reduces to $\hat \Phi_{i,j} = 1 $ for $j=i+1$. 
Then, applying the JW transformation to Eq.~(\ref{eq:H}) we obtain (neglecting irrelevant boundary terms):
\begin{align}
 H^F =\sum_{\beta=1,2} \frac{J_\beta}{2}\sum_{i=1}^L \left( c^\dagger_i c_{i+\beta} + \text{H.c} \right) + V_{\text{int}}+\sum_ih_i\hat n_i,
 \label{eq:Hf}
\end{align}    
with the non-local interaction (correlated hopping) term 
\begin{align}
V_{\text{int}}=-J_2\sum_{i=1}^L \left( c^\dagger_i \hat{n}_{i+1} c_{i+2} + \text{H.c} \right).
\label{eq:Int}
\end{align}

We thus have an interacting fermionic system  with a non-local interaction in a potential disorder. In the absence
of frustration (i.e. for $J_2 = 0$) the interaction and the next-nearest-neighbor hopping terms vanish, and  in Eq.~(\ref{eq:Hf}) one has a system of free fermions with nearest-neighbor
hopping in the potential disorder, where all states are localized~\cite{Mott,Klein}. It is the non-local interaction $V_{\text{int}}$ that can lead to delocalization at weak disorder. This becomes more evident if one replaces the first term in the Hamiltonian (\ref{eq:H}) for $\beta=2$ with  $ J_2 \sum_{i} \left( S^x_iS^x_{i+2} +S^y_iS^y_{i+2} \right)  S_{i+1}^{z}$.
We then arrive at the so-called XZX+YZY model which maps onto  the free-fermionic limit of Eq.~(\ref{eq:Hf})~\cite{XZXYZY}, where $V_{\text{int}}=0$ but there is the next-nearest-neighbor
hopping on top of the nearest neighbor one. The states in this case are localized in a weak disorder~\cite{Mott}.      
 We note that the longer-range flip-flop term in Eq.~(\ref{eq:H}) also fermionizes onto many-body interaction terms encapsulated in the JW phase Eq.~(\ref{eq:Phase}). In general, the flip-flop transition to the $\beta$ nearest sites generates up to $\beta$-body interaction. This can be seen already in the $\beta=3$ case: 
	\begin{align}
	H_3  = J_3 c^\dagger_i (\hat{1}-2\hat{n}_{i+1}-2\hat{n}_{i+2}+4\hat{n}_{i+1}\hat{n}_{i+2})c_{i+3} .
	\end{align}
Here, the first term corresponds to the trivial fermionic hopping, while the other terms imply many-body interactions. Thus, one has to take into account this type of interaction terms in the studies of frustrated spin models~\cite{KhemaniHuse}.

{\it{Mapping onto the system of hard-core bosons.}} A simple qualitative explanation of the present interaction can also be given using alternative representation of the clean model~(\ref{eq:H}) in terms of hard-core bosons  via the Matsubara-Matsuda transformation~\cite{MatsubaraMatsuda},
 \begin{equation}
 S^+_i=b^\dagger_i, \quad  S^z_i=b^\dagger_ib_i-\frac{1}{2},
 \end{equation}
\begin{equation}
H^B=\sum_{i,\beta=(1,2)} \frac{J_\beta}{2}(b^\dagger_ib_{i+\beta} + \text{H.c} ).
\label{HB}
\end{equation} 
The Hamiltonian~(\ref{HB}) consists of only the kinetic term for bosons, with an imposed constraint $b^\dagger_i b_i\leq 1$ for a given site $i$. In general, such a hard-core constraint leads to a {\it{hard-core interaction}}. The origin of this interaction lies in the truncation of the normal bosonic Hilbert space by the constraint on real-space occupations. This interaction does not manifest itself in~(\ref{HB}) for $J_2=0$ (or $J_1=0$), when the geometry reduces to the strict 1D chain. In this case, hard-core bosons hop around the ring, while strictly keeping their ordering, i.e. no particle exchange occurs. It is this ordering that guarantees JW mapping onto free spinless fermions, where the hard-core constraint plays a role of the Pauli principle. 
At finite $J_2$, the single-particle behavior does not hold since the additional hopping channel is introduced and particle exchange is  no longer prohibited. The latter guarantees the manifestation of hard-core interaction~\cite{Spin1Note}, which is exhibited in the form of (\ref{eq:Int}) in the fermionic formulation. We note that the interplay of disorder and hard-core interaction in quasi-1D geometries was already addressed previously in the context of superfluid-Bose glass transition at zero temperature \cite{BoseGlass1,BoseGlass2}.

     Presented analytical arguments and qualitative discussions hint at the MBL transition when the disorder strength is gradually increased. In the next sections, we numerically examine this prediction based on finite-size ED calculations.

\section{Characterization of the MBL transition} 
 An important quantity that characterizes the eigenstates of the disordered model is the ratio of the minimum to maximum consecutive level spacing,
\begin{equation} \label{eq:r}
 r_i =\frac{\min (\Delta_i,\Delta_{i+1})}{\max (\Delta_i,\Delta_{i+1})},
\qquad
\Delta_i = \epsilon_i-\epsilon_{i-1},
\end{equation}
where $\epsilon_i$ are the ordered energy eigenvalues for a given realization of disorder. In the delocalized (chaotic) phase the energy level spacing distribution obeys Wigner's surmise of the Gaussian orthogonal ensemble (GOE), while in the localized phase no level repulsion is expected and there is a Poissonian distribution (PS) of the level spacings. For the PS distribution the disorder-averaged value is $\langle r\rangle_{\rm P}=2\ln 2-1 \approx 0.386$, and for the Wigner-Dyson (WD) distribution one has $\langle r\rangle_{\rm W} = 0.5307(1)$ \citep{OganesyanHuseRn}.
 
 We also consider localization of eigenstates in the Hilbert space. For the model~(\ref{eq:H}) with $L$ sites and a fixed number of $\lbrace \uparrow  \rbrace$ spins $M$ one has ${\cal N_H}$-dimensional Hilbert space, with ${\cal N_H}={\cal{C}}^L_{M}=\frac{L!}{M!(L-M)! }$.
  We analyze many-body eigenstates in the computational basis $|s \rangle= |s_1 \rangle \otimes |s_2\rangle \otimes ... \otimes |s_L \rangle$, with local states $|s_i\rangle \in \left\lbrace  |\uparrow  \rangle, |\downarrow \rangle \right\rbrace$. 
The quantities well characterizing localization properties of the wavefunctions 
$\psi_\alpha(s)=\langle s  |  \alpha  \rangle $ are the fractal dimensions $D_q$. The set of $D_q$ is determined from the scaling of participation entropies $S_q$ with~${\cal N_H}$,
\begin{equation}
S_q=\frac{1}{q-1}\ln\left(\sum_{s=1}^{\cal N_H} |\psi_\alpha(s)|^{2q}\right) \xrightarrow{{\cal N_H} \rightarrow \infty} D_q \ln\left({\cal N_H}\right) .
\label{eq:Dq}
\end{equation}
Eigenstates $|\alpha \rangle$ localized on a finite set of $|s \rangle$ have $S_q$ independent of ${\cal N_H}$ and thus $D_q=0$ for any $q>0$. On the other hand, delocalized states with $|\psi_\alpha(s)|^2 \sim {\cal N_H}^{-1} $ give $D_q=1$. The multifractal states with $0<D_q<1$ are non-ergodic albeit extended. We confine ourselves to the \textit{Shannon limit} ($q\rightarrow1$) in~Eq.~(\ref{eq:Dq}).
 
Quantum correlations between neighboring (in energy) eigenstates can be calculated directly if the eigenstates are known and hence can also be used to characterize the MBL transition. The corresponding quantity is the Kullback-Leibler divergence $KL$~\cite{Kullback1951,KravtsovCit,Pino2019}:
\begin{equation} 
KL=\sum_{s=1}^{\cal N_H} | \psi_{\alpha}(s) |^2 \ln \left( \frac{|\psi_{\alpha}(s)|^2}{|\psi_{\alpha+1}(s)|^2} \right),
\label{eq:KL}
\end{equation}  
where the states $|\alpha \rangle$  are supposed to be ordered in energy. The extended states close to this transition can be viewed as the result of the hybridization of localized states. The localized states are not correlated in space and the ratio $| \frac{\psi_{\alpha}(s)}{\psi_{\alpha+1}(s) }|$ is exponentially large if $|\psi_{\alpha}(s)|$ is not negligible, i.e. $KL\rightarrow \infty$ for ${\cal N_H} \rightarrow \infty$. After the hybridization, extended states $|\alpha \rangle$ and $|\alpha+1 \rangle$ involve mostly the same localized states. As a result $|\psi_{\alpha}(s)|$ and $|\psi_{\alpha+1}(s)|$ are strongly correlated with $| \frac{\psi_{\alpha}(s)}{\psi_{\alpha+1}(s) }| \sim  O(1)$ and $KL$ is finite. An abrupt change of $KL$ is therefore an indication of the MBL transition.

We study the MBL transition for eigenstates with energies close to zero (eigenstates from the central part of the spectrum), although this transition can be observed at any energy density (assuming sufficiently strong frustration to delocalize low-energy eigenstates). 
For lattice sizes $L=\left\lbrace  14,16,18,20 \right\rbrace$ we employ the shift-invert ED algorithm based on $\mathsf{LDL^T}$ decomposition to obtain $m=\left\lbrace 20,20,40,100 \right\rbrace$ eigenstates. We consider the problem in the largest Hilbert subspace (at total magnetization ${\cal S}^z=0$) with dimensions ${\cal N_H}=\binom{L}{L/2}$. However, we note that the MBL transition in the other sectors also can be observed. The number of disorder realizations for a given disorder strength~$h$ varies from $10^4$ for the smallest lattice size up to $10^2$ for the largest size. 
 We average the quantities of our interest over the ensemble of $m$ states and then the disorder averaging is performed.
 
  \label{sec:numerics}
\section{Numerical results}
\begin{figure}[t]
\includegraphics[width=8.5cm ]{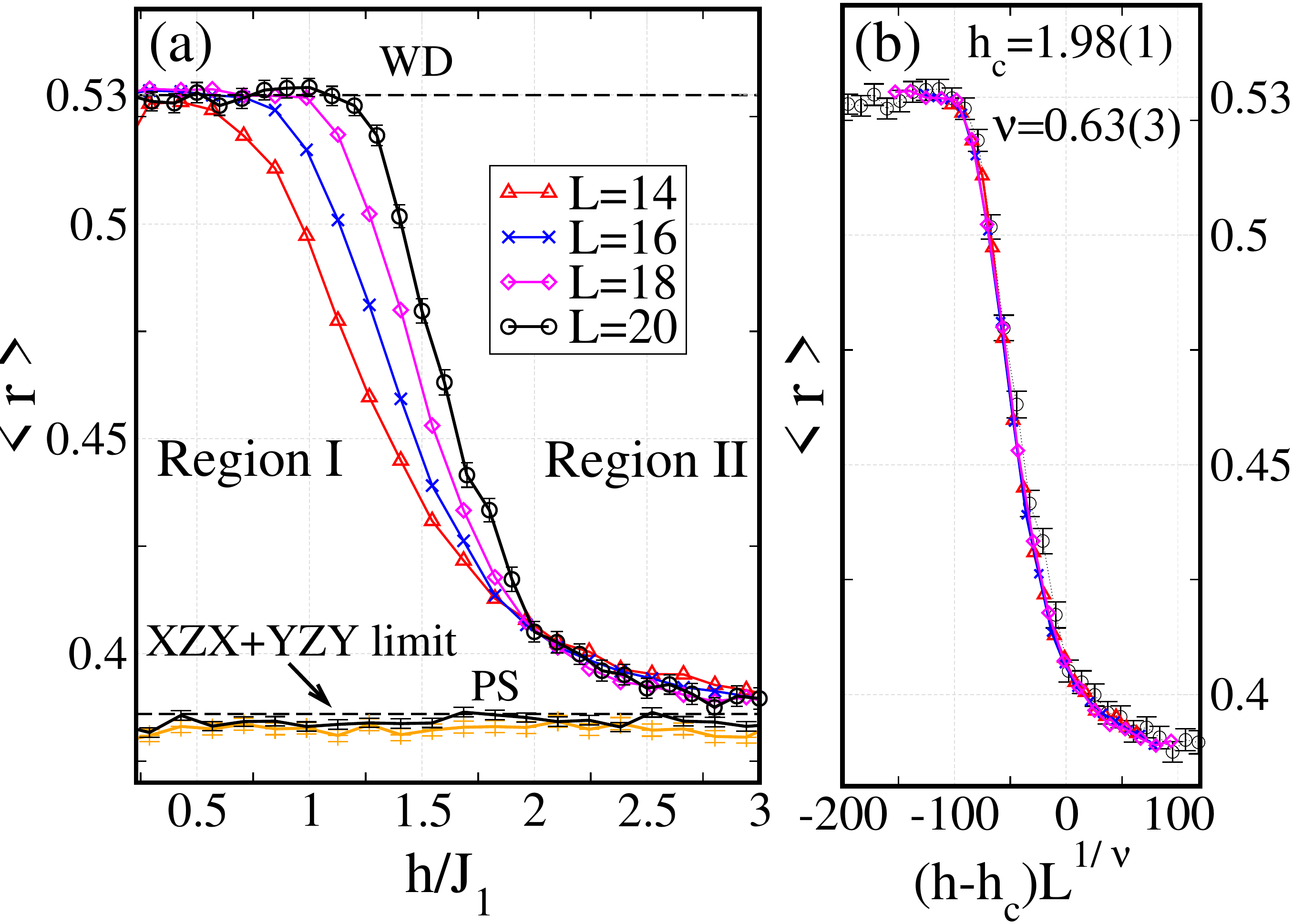}  
 \caption{\label{fig:fig1} 
 (a) Dependence of the average gap ratio $\langle r \rangle$ on the disorder strength $h$ for the system sizes $L= \left\lbrace 14,16,18,20 \right\rbrace $ at fixed $\kappa=0.1$.
  The MBL phase (Region II) is separated from the delocalized phase (Region I)  at the critical field amplitude $h_c/J_1 \approx  1.98 $. In the XZX+YZY limit the states are localized and $ \langle r \rangle = \langle r\rangle_{\rm p} \approx 0.386$ at arbitrary disorder strength. 
 (b) Implementation of the scaling ansatz leads to the collapse of numerical data to a single universal curve with $h_c/J_1 =  1.98\pm 0.01$ and $\nu=0.63\pm 0.03$.    
} 
\end{figure}

 In this section we present our numerical results at fixed $\kappa=0.1 $. All calculated quantities confirm the presence of a thermal delocalized phase at weak disorder, whereas at large disorder the system undergoes the MBL transition. We firstly identify the MBL transition exploiting the average gap ratio for adjacent eigenstates $\langle r\rangle$, defined in Eq.~(\ref{eq:r}). Our findings are illustrated in Fig.~\ref{fig:fig1}(a). 
 
For the disorder strength $h/J_1 \lesssim 1$ the energy mini-gaps obey the WD statistics with $\langle r\rangle_{\rm W} \approx 0.53$. This  implies the presence of hybridization between regions of the system, which results in the level repulsion between the neighboring eigenstates. For all system sizes that we consider the benchmarked WD level $\langle r\rangle_{\rm W}$ is observed in a finite interval of $h/J_1$, although for our smallest system size $L=14$ the finite-size effects are the strongest and the WD distribution is not fully obeyed. When the disorder strength is further increased, the energy mini-gap statistics deviates from the WD distribution taking the full PS chatacter at large disorder. 
The curves corresponding to different $L$ cross each other in the vicinity of the critical point $h/J_1 \approx 2$. When the system size is further increased the crossing points drift to larger values of disorder strength and in the thermodynamic limit the convergence of the crossing points to the critical point is expected. Based on the finite-size calculation results, the behavior of $ \langle r \rangle$ near the critical point can be analyzed via the scaling form $\langle r \rangle \sim  f\left( L^{1/\nu}(h-h_c) \right) $, with the scaling function $f$ \cite{LuitzAletRn}. Direct implementation of this scaling ansatz leads to the collapse of all data to a single universal curve (see Fig.~\ref{fig:fig1}(b)) with $h_c/J_1 = 1.98 \pm 0.01$ and $\nu = 0.63\pm 0.03$. 
 For the disorder strength  $h > h_c $ the states are in the MBL phase and the gap statistics converges to the PS distribution with $\langle r\rangle_{\rm P} \approx 0.39$.    
    
The character of the demonstrated MBL transition is similar to the one for the 1D Heisenberg chain in a random magnetic field~\cite{LuitzAletRn,OganesyanHuseRn}. The ergodic phase at weak disorder in the latter case is guaranteed by the Ising interaction term ($\hat{n}_i\hat{n}_{i+1}$ for dual fermions), whereas in our model it appears due to the interaction term $V_{\text{int}}$. 
To finalize this argument, we repeated our calculations of  $\langle r \rangle$ for the system sizes  $L=\left\lbrace 16, 18 \right\rbrace$ when the interaction term is removed (XZX+YZY limit). The corresponding plots are presented in Fig.~\ref{fig:fig1}(a). For both considered system sizes a stable PS distribution of level spacings with $\langle r \rangle = \langle r \rangle_{\rm p} \approx 0.39$ is exhibited, which corresponds to Anderson localization of eigenstates.      

 The MBL transition is also correctly captured by the fractal dimensions presented in Fig.~\ref{fig:fig2}. In the delocalized phase the support set of  states covers a sufficiently large fraction of the Hilbert space with $|\psi(j)|^2 \sim {\cal N_H}^{-1} $ in the thermodynamic limit. This implies $D_q \rightarrow 1 $, $S_q \sim \ln({\cal N_H})$ as ${\cal N_H} \rightarrow \infty$. In our finite-size calculations $\langle D_1 \rangle$ is practically independent of $h/J_1$ in a finite interval of $h/J_1$ with the values $\langle D_1\rangle=\left\lbrace 0.900, 0.922, 0.931, 0.939 \right\rbrace$ for the system sizes $L=\left\lbrace 14, 16, 18, 20 \right\rbrace$, respectively.  The convergence of~$\langle D_1 \rangle$ towards unity with increasing the system size is demonstrated clearly by the histograms with a vanishing variance~Fig.~\ref{fig:fig2}(a). In the thermodynamic limit the constant value $ D_1=1$ is expected within the chaotic phase, with an abrupt jump to $ D_1=0$ at the critical field. In the MBL phase the distribution of $D_1$ has an opposite skewness and converges towards zero with growing the system size  as shown in~Fig.~\ref{fig:fig2}(b). As expected, the benchmarked critical point $h_c/J_1 = 1.98(1)$ lies within the transition area (shaded region) determined from $D_1$ curve-crossings.
 
\begin{figure}[t]
\includegraphics[width=8.5cm]{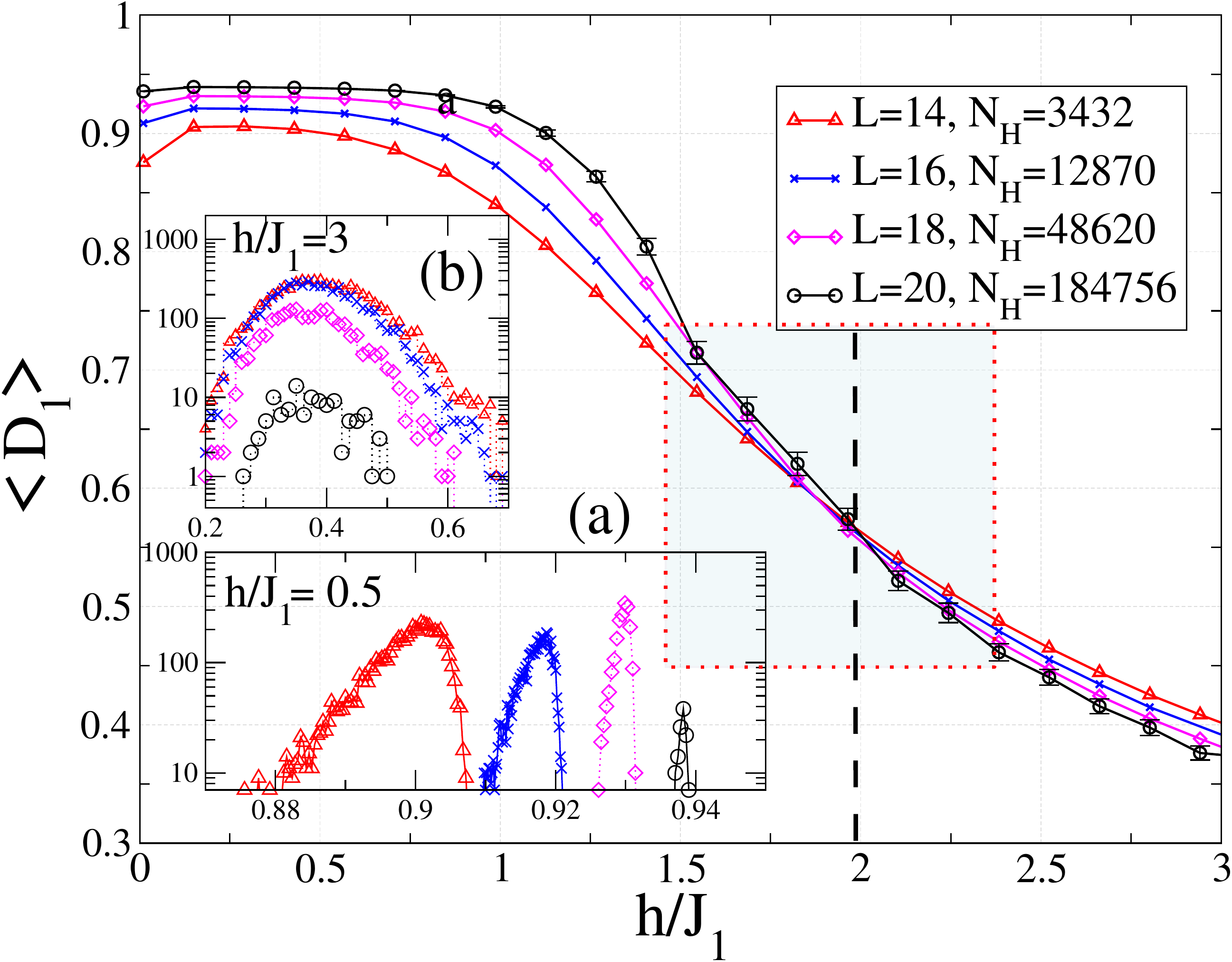}%
\label{fig:fig2} 
\caption{\label{fig:fig2} 
 Averaged fractal dimensions $ \langle D_1 \rangle$  versus the disorder strength~$h/J_1$  for $L= \left\lbrace  14,16,18,20 \right\rbrace $. The critical point $h_c/J_1 = 1.98(1)$ is fixed by the vertical dashed line and correctly lies in the shaded region, defined by the $\langle D_1 \rangle$ curve-crossings. Insets show scaled histograms $P(D_1)$ for two representative disorder amplitudes (a)  at the delocalized phase ($h/J_1=0.5$) and (b) at the MBL phase ($h/J_1=3$). Error bars smaller then the symbol size are omitted.
}
\end{figure}
 

 Strong quantum correlations between nearby eigenstates in the chaotic regime result in the known value of  $\langle KL \rangle =  KL_{\text{GOE}}=2$, which is demonstrated in Fig.~\ref{fig:fig3} (left). This value is kept within the delocalized region, which widens with increasing the system size. On the other hand, neighboring eigenstates in the MBL phase are weakly correlated and this results in the divergent behavior $\langle KL \rangle \sim \ln(\cal{ N_H})$.   These distinctive features are also demonstrated in the scaled histogram plots in the insets of Fig.~\ref{fig:fig3}. In the delocalized phase [Fig.~\ref{fig:fig3} (a)] the distribution of $KL$ has  a Gaussian form with the mean value $ \langle KL \rangle=2$ and  the variance vanishing with increasing $L$. On the contrary, in the MBL phase the the situation is different and both the mean value and the variance increase with $L$ [Fig.~\ref{fig:fig3} (b)]. Although a large drift of crossing points with increasing the system size does not allow one to do a precise finite-size scaling analysis, it is clear that the crossing point of the last two curves (corresponding to the largest Hilbert spaces) lies in the critical region determined above from $D_1$ and it is close to the benchmarked critical field $h_c$.

 We next performed ED calculations of $D_1$ for other values of $\kappa = J_2 / J_1 \lesssim 0.32$, such that the system is still in the gapless phase. The determined finite-size critical field strengths $h_c(\kappa)/J_1$ based on curve-crossings of $D_1$    (corresponding to $L=14$ and $L=16$) are presented in  Fig.~\ref{fig:fig3} (right).  This figure demonstrates that an increase of $\kappa$ should increase the interaction strength in Eq.(\ref{eq:Int}) and, hence, should lead to the enhancement of the delocalization effect of the frustration. This indeed results in a linear growth of $h_c/J_1$ with $\kappa$, as shown in Fig.~\ref{fig:fig3} (right). In the limit of $\kappa \rightarrow \infty$ we arrive at two weakly coupled $XY$ chains and the critical field should decrease, reaching $h_c=0$ at $\kappa=\infty$.

 
\begin{figure}[t]
\includegraphics[width=8.5 cm ]{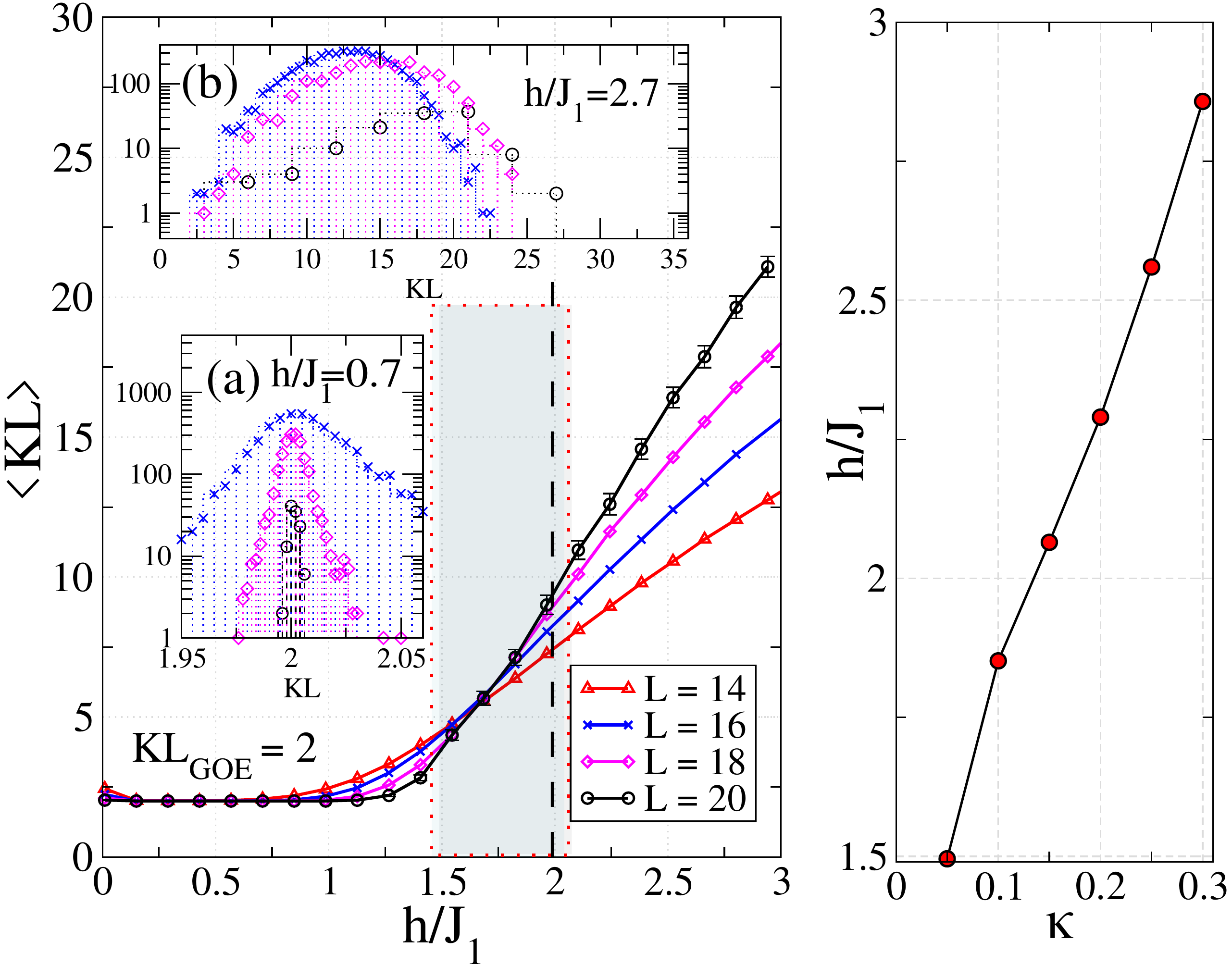}%
\label{fig:fig3} 
\caption{\label{fig:fig3} 
 The Kullback-Leibler divergence $\langle KL \rangle$ as a function of disorder strength~$h/J_1$  for $L= \left\lbrace  14,16,18,20 \right\rbrace $ (left) and finite-size phase boundary in the  $\kappa - h $ space (right). Insets show scaled histograms $P(KL)$ (a) at the MBL phase ($h/J_1=2.7$) and (b) at the delocalized phase ($h/J_1=0.7$). In the left pannel, the critical field is fixed by the vertical dashed line and it lies in the shaded region, defined by the $\langle KL \rangle$ curve-crossings.   
}
\end{figure}

\label{sec:discussion}


\section{Outlook and conclusions}
 Actually, the investigation of localization-delocalization transitions in disordered systems with long-range interactions and long-range hopping terms has a fairly long history. The first consideration of the long-range hopping terms was done in the seminal work of Anderson on disordered $ d $-dimensional non-interacting fermionic systems with $ J_{ij} \sim r_{ij}^{-\alpha} $ ($J_{ij}$ and $r_{ij}$ are the hopping amplitude and distance between the $i$-th and $j$-th sites, respectively). It predicts delocalization for $\alpha  \le d $ at any disorder strength even at~$T=0$~\cite{Anderson}. A natural extension of this study to many-body quantum systems with the Ising-type long-range interaction $V_{ij} \sim  r_{ij}^{-\alpha} S^z_iS^{z}_{j}$, modifies the above criterion to $\alpha <2d$ \cite{BurinKagan,Lauman,BurinXYZ}, implying a possibility of the MBL transition for $\alpha > 2d$.  
 
 The natural extension of our model for longer-range flip-flop amplitudes was addressed in $d$-dimensional case with $J_{ij} \sim r_{ij} ^{-\alpha}$ in a random transverse magnetic field~\cite{BurinXY}. It was shown that being explicitly absent initially, the Ising-type interaction is generated in the third order of perturbation theory in hopping~\cite{BurinXY}. This interaction delocalizes states for $ \alpha > 3d/2 $ provided that the disorder is sufficiently weak. Numerical findings based on finite-size ED ($L \leq 18 $) and matrix product states ($L \leq 40$) suggest the  MBL transition for $\alpha   \gtrsim 1 $ in 1D systems~\cite{MPSstudy,EDMBLXY}. Experimentally, the 1D XY chain with long-range flip-flop amplitudes was realized using trapped Ca$^+$ ions (with $\alpha<3$) \cite{ExpLRHXY}.

We have demonstrated that in a 1D XY magnet in a random magnetic field already a weak frustration provided by next-nearest-neighbor hopping is sufficient for the emergence of many-body localization-delocalization transitions. We exploited the Jordan-Wigner transformation to map the spin model onto a system of interacting fermions with a non-local interaction originating from the frustration. Our results for the level-spacing statistics show the presence of quantum correlations between the neighboring  eigenstates and the resulting level repulsion in the delocalized phase. In the limit of strong disorder, the neighboring eigenstates are not correlated and the system is in the MBL phase. This leads to the vanishing level repulsion, and energy mini-gaps obey the PS statistics. The present results are supported by the calculation of fractal dimensions and characterization of quantum correlations of neighboring eigenstates by the KL divergence. 
	
 While our primary aim in this work is the demonstration of the MBL transition, there are open questions to be addressed. Particularly, the nature of the high-temperature spin transport in the delocalized phase remains to be investigated. This problem was studied previously for the disordered 1D XXZ model and the low-frequency optical conductivity of the form $\Re\sigma(\omega)=\sigma_{dc}+c|\omega|^\delta$ with $\sigma_{dc}>0$ and $\delta \sim 1$ was proposed~\cite{RobinSt,Barisic2016,PreLov}. However, in our model the effect of the frustrating term on transport properties is expected to be drastic. In the disorder-free case, this term degrades integrability blocking ballistic spin transport. Such frustrating perturbation of the 1D Heisenberg chain is known to cause a crossover from subdiffusive to superdiffusive transport, depending on the sign of $J_2$~\cite{AnTransport}. This puts the applicability of the proposed form with $\delta \sim 1 $ for our model with finite disorder into question. Thus, transport properties of the model $(\ref{eq:H})$ in the clean and disordered cases should be separately studied in the future.   
 
In summary, our results demonstrate an important role of frustration terms in disordered spin models. The discussed transition can be equally observed in other type of quasi-1D ladders, where the JW phase survives and generates a many-body interaction. Experimentally, the discussed MBL transition can be realized, e.g., by using a setup of superconducting qubit arrays proposed in Ref.~\cite{Proposal}.

\begin{acknowledgments}
We thank V. Gritsev and W. Buijsman for useful comments and discussions. This research was supported in part through computational resources of HPC facilities at HSE University \cite{HPCHSE} and by the Russian Science Foundation Grant No. 20-42-05002. We also acknowledge support of this work by Rosatom.
\end{acknowledgments}

\appendix 


  \section{Exact form of the conserved charges for the 1D XY model }

 We firstly consider conserved quantities (charges) for the spin-$1/2$ isotropic XY-model in an inhomogeneous magnetic field. The Hamiltonian is given by
\be \label{H_XXh_spins}
	H_0 = J_1 \sum_j \left( S_j^x S_{j+1}^x + S_j^y S_{j+1}^y  \right) + \sum_j h_j S_j^z,
\ee
where~$S_j^{\alpha} = \sigma_j^{\alpha}/2$ with $\alpha \in \{ x, y, z\}$, and $\sigma_j^{\alpha}$ are the Pauli matrices acting non-trivially on the $j$-th site of the chain, $J_1$ is the coupling constant for the nearest-neighbor exchange interaction between the spins, and~$h_j$ is the inhomogeneous transverse magnetic field.
The Hamiltonian~(\ref{H_XXh_spins}) is integrable for any boundary condition and with arbitrary~$h_j$. It can be diagonalized by the Jordan-Wigner transformation, which reduces Eq.~(\ref{H_XXh_spins}) to the model of non-interacting spinless fermions~\cite{S_Lieb1961}.

In the homogeneous case, i.e.~$h_j = h$, there are two families of conserved charges, which commute with~$H_0$ and each other. Explicitly, they are given by~\cite{S_Gusev1982, S_Grabowski1995}
\be \label{XXh_homogeneous_Q}
 \begin{aligned}
 	Q_{n}^{(1)} &= \sum_j \left( e_{n, j}^{x y} - e_{n, j}^{y x} \right), \\
	Q_{n}^{(2)} & = J_1 \sum_j \left( e_{n, j}^{x x} + e_{n, j}^{y y} + e_{n-2, j}^{x x} + e_{n-2, j}^{y y} \right)\\
	&-h \sum_{j} \left( e_{n-1, j}^{x x} + e_{n-1, j}^{y y} \right),
 \end{aligned}
\ee
where by convention~$n \geq 3$ and we denoted
\be
	e_{n,j}^{\alpha, \beta} = S_j^{\alpha} S^z_{j+1} \ldots S^z_{j+n-2} S_{j+n-1}^{\beta}, \qquad e_{1,j}^{\alpha \alpha} = - \, S_j^z.
\ee
Note that the combinations~$\sum_j \left( e_{n,j}^{xx} + e_{n,j}^{yy} \right)$ and $\sum_j \left( e_{n,j}^{xy} - e_{n,j}^{yx} \right)$ commute with the total magnetization~$S^z = \sum_{j} S_j^z$.
 
We now turn to the case of an inhomogeneous magnetic field, i.e. $h_j$ is an {\it arbitrary} function of the lattice site~$j$. One can show that the Hamiltonian~(\ref{H_XXh_spins}) commutes with the following conserved charges
\be \label{XXh_inhomogeneous_Q}
	{\cal Q}_n = \sum_j \sum_{k=0}^{n-2} a_j^{(k)} \left( e_{n-k, j }^{x x} + e_{n-k, j }^{y y} \right) - \sum_j a_j^{(n-1)} S_j^{z},
\ee
given that the coefficients~$a_j^{(m)}$ in Eq.~(\ref{XXh_inhomogeneous_Q}) satisfy a set of recurrent relations:
\be \label{XXh_Q_coeffs_relations}
\begin{aligned}
  &J_1 \left( a^{(m)}_{j+1} - a_j^{(m)} \right) = J_1 \left( a^{(m-2)}_{j} - a_{j-1}^{(m-2)} \right)\\  
 &- a_j^{(m-1)} \bigl( h_{j+n - m} - h_j \bigr),\quad  \left( 0 \leq m \leq n-1\right),
 \end{aligned}
\ee
where for $l<0$ we have~$a_j^{(l)} \equiv 0$. Then, taking $m=0$ in Eq.~(\ref{XXh_Q_coeffs_relations}) we immediately see that~$a_{j+1}^{(0)} = a_j^{(0)}$, which has only a homogeneous solution $a_j^{(0)} = a_1^{(0)}$. The rest of $n-1$ equations in Eq.~(\ref{XXh_Q_coeffs_relations}) can be successively solved to determine~$n-1$ remaining coefficients. Thus, we obtain
\be
a_j^{(m)} = a^{(m)}_1 + A_j^{(m)},
\ee
\be
\resizebox{0.48\textwidth}{!}{$A_j^{(m)}=\sum_{k=1}^{j-1} \left[ a_k^{(m-2)} - a_{k-1}^{(m-2)} - \frac{1}{J_1} \, a_k^{(m-1)} \left( h_{k+n-m} - h_k \right) \right]$},
\ee
which is valid for arbitrary~$h_j$. It is easy to check that conserved charges~${\cal Q}_n$ in Eq.~(\ref{XXh_inhomogeneous_Q}) commute with each other. Clearly, for the homogeneous case,~$h_j = h$, the charges~(\ref{XXh_inhomogeneous_Q}) coincide with~$Q_{n}^{(2)}$ in Eq.~(\ref{XXh_homogeneous_Q}), i.e. only one of the families of conserved charges survives in the presence of the inhomogeneous field.

 \section{Quasiconserved charges in the perturbed XY model with the homogeneous field}
We next consider the Hamiltonian~$H = H_0 + H_1$, where $H_0$ is the integrable part given by Eq.~(\ref{H_XXh_spins}) with the homogeneous magnetic field, $h_j = h$, and $H_1$ is a perturbation that breaks integrability. The perturbation in our case has the following form:
\be \label{H_1_spins}
	H_1 = J_2 \sum_j \left( S_j^x S_{j+2}^x + S_j^y S_{j+2}^y  \right).
\ee

We assume that the perturbation is weak and one has~$\kappa = J_2/ J_1 \ll 1$. It is believed that in the case of weak integrability-breaking perturbation the model is quasi-integrable. In particular, this implies that it should not thermalize for times as large as~$\tau_{\text{th}} \sim \kappa^{-2}$, so that the model possesses {\it approximate} conservation laws that prevent thermalization at shorter times~\cite{S_prethermalization, S_prethernearin1, S_prethermnearin2, S_pretherml1, S_pretherml2}. 

Clearly, in the presence of the perturbation~(\ref{H_1_spins}) the charges $Q_n^{(1,2)}$ from Eq.~(\ref{XXh_homogeneous_Q}) are no longer conserved, since they do not commute with the term~$H_1$. They are not even quasiconserved, because one has~$|| [ H_1, Q_n^{(1,2)} ]  || \propto \kappa$.  Therefore, under the evolution with the Hamiltonian~$H = H_0 + H_1$ the operators~$Q_n^{(1,2)}$ change significantly at times much shorter than~$\tau_{\text{th}} \sim \kappa^{-2}$ and can not be responsible for the non-ergodic behaviour in the prethermal phase. Using the procedure discussed in detail in Ref.~\cite{Kurlov2021}, one can show that the first non-trivial quasiconserved charge reads
\be
	\tilde Q_3 = Q_3^{(1)} + \kappa \, \delta Q_3,
\ee
where~$Q_3^{(1)}$ follows from Eq.~(\ref{XXh_homogeneous_Q})  with~$n=3$ and the correction~$\delta Q_3$ is given by
\be
  \begin{split}
	&\delta Q_3 =  \sum_j S_j^x S_{j+2}^z S_{j+3}^y  - \sum _j S_j^y S_{j+2}^z S_{j+3}^x   \\
	& +\sum_j S_j^x S_{j+1}^z S_{j+3}^y  -\sum _j S_j^y S_{j+1}^z S_{j+3}^x  \\
	& +\sum_j \left( {\boldsymbol S}_j \times {\boldsymbol S}_{j+1} \right) \cdot {\boldsymbol S}_{j+2},
\end{split}
 \ee
 where ${\boldsymbol S}_j = \{ S_j^x, S_j^y, S_j^z  \}$ is the vector of spin operators in the $j$-th site. One can easily check that $\tilde Q_3$ satisfies the relation 
 \be
 	\left\lVert \left[ H_0 + H_1, \tilde Q_3 \right] \right\rVert_F \propto \kappa^2,
 \ee
 where $|| X ||_F = \sqrt{\text{tr }X^{\dag} X}$ is the Frobenius norm. One can also obtain higher-order quasiconserved charges~$\tilde Q_n$, which commute with the Hamiltonian~$H_0 + H_1$ and each other with the accuracy~${\cal O}(\kappa^2)$. However, this is beyond the scope of the present paper.



\begin{thebibliography}{99}
\bibitem{XY}
	 We use the name 'XY-model' following quantum magnetism community. However, it is also often called the XX-model, since the exchange interaction is isotropic in the XY-plane.
\bibitem{Spin1Note}
One can alternatively achive the manifestation of the hard-core interaction in the model~(\ref{HB}) with $J_2=0$ by increasing the number of local degrees of freedom in each site. The trivial case is to allow $2$ bosons to occupy a given site, i.e $b^\dagger_i b_i\leq 2$. The resulting model is dual to the spin-1 XY model, which also exhibits the MBL transition in a random magnetic field in the $z$ direction. This transition will be discussed elsewhere.
\end{thebibliography}

  \end{document}